\documentclass{article}
\usepackage[T1]{fontenc}
\usepackage[utf8]{inputenc}
\usepackage{ismir} 
\usepackage{amsmath,cite,url}
\usepackage{graphicx}
\usepackage{color}

\usepackage{xcolor}
\usepackage[dvipsnames]{xcolor}
\usepackage[table]{xcolor}
\usepackage{amssymb}
\usepackage{booktabs}
\usepackage{pifont}
\usepackage{enumitem}

\definecolor{color_a}{RGB}{195,164,207}
\definecolor{color_b}{RGB}{229,212,232}
\definecolor{color_c}{RGB}{249,249,249}
\definecolor{color_d}{RGB}{217,241,213}
\definecolor{color_e}{RGB}{173,212,160}

\title{Integrating Contextual Embeddings into Evaluation of Expressive MIDI Piano Performances}

 \multauthor
   {Dmitrii Gavrilev$^1$ \hspace{1cm} Ilya Borovik$^1$ \hspace{1cm} Vladimir Viro$^2$}
   {
   $^1$ Skolkovo Institute of Science and Technology, Russia\\
   $^2$ Peachnote GmbH, Germany\\
   {\tt\small dmitrii.gavrilev@skoltech.ru}
   }

\def\authorname{D. Gavrilev, I. Borovik, and V. Viro}

\usepackage[bookmarks=false,pdfauthor={\authorname},pdfsubject={\pdfsubject},hidelinks]{hyperref}

\begin{document}

\maketitle

\begin{abstract}
Objective evaluation of expressive MIDI piano performances typically relies on attribute statistics such as timing, velocity, and duration of individual notes. However, these methods often disregard dependencies between notes, which poses a potential limitation in assessing the similarity between two sets of performances. In generative applications, the wide variety of expressive attributes makes it difficult to aggregate them into a single scalar metric for model selection. In this work, we reexamine attribute-scoped metrics and explore the perceptual properties of contextual embeddings from self-supervised symbolic music models, Aria and CLaMP3. Results from our listening study indicate that these models can be used as perceptual proxies, showing agreement with per-sample human ratings on par with traditional metrics. To measure conditional distributional similarity, we adapt Kernel Audio Distance to the symbolic music domain. Unlike Pearson correlation and reconstruction error, kernel-based methods on contextual embeddings do not require note alignment and are sensitive to contextual perturbations. To facilitate reproducibility, we release Pereval\footnote{Available at \href{https://github.com/realfolkcode/pereval/}{https://github.com/realfolkcode/pereval/}}, an open-source library that integrates performance evaluation utilities, including both attribute-scoped and deep feature metrics.
\end{abstract}

\section{Introduction}\label{sec:intro}
In recent years, research in Music Information Retrieval (MIR) has shown growing interest in understanding and modeling the expressive aspects of music. A significant branch of MIR focuses on processing symbolic music using formats such as MusicXML and MIDI. The latter allows researchers to isolate the expression attributes of individual notes, albeit at the expense of disregarding acoustic features \cite{cancino2017evaluation}. However, despite the growing interest, performance-level assessment of music remains underexplored \cite{Zhang2025LLaQo}.

Common examples of expressive attributes include tempo, loudness, and articulation \cite{cancino2018computational}. These attributes are often used to compare two sets of performances, e.g., human and generated corpora of MIDI samples. Traditional metrics, which we coin attribute-scoped, rely on comparisons between individual attributes, i.e., they can "hear" only one expressive facet at a time \cite{peter2023sounding}. Another limiting factor is that most commonly adopted metrics either require fine structural alignment (e.g., note-wise), which hinders the scalability of evaluation, or aggregate the note statistics independently of sheet music.

\begin{table}[t]
\centering
\resizebox{0.45\textwidth}{!}{%
\begin{tabular}{@{}lcccccc@{}}
\toprule
 & Fidelity & Diversity & Alignment-free & Score-aware & Contextual & Per-sample \\
\midrule
Reconstruction error & \textcolor{ForestGreen}{\ding{52}} & \textcolor{red}{\ding{56}} & \textcolor{red}{\ding{56}} & \textcolor{ForestGreen}{\ding{52}} & \textcolor{red}{\ding{56}} & \textcolor{ForestGreen}{\ding{52}} \\
Inter-correlation & \textcolor{ForestGreen}{\ding{52}} & \textcolor{red}{\ding{56}} & \textcolor{red}{\ding{56}} & \textcolor{ForestGreen}{\ding{52}} & \textcolor{red}{\ding{56}} & \textcolor{ForestGreen}{\ding{52}} \\
Intra-correlation & \textcolor{red}{\ding{56}} & \textcolor{ForestGreen}{\ding{52}} & \textcolor{red}{\ding{56}} & \textcolor{ForestGreen}{\ding{52}} & \textcolor{red}{\ding{56}} & \textcolor{red}{\ding{56}} \\
KL Divergence & \textcolor{ForestGreen}{\ding{52}} & \textcolor{ForestGreen}{\ding{52}} & \textcolor{ForestGreen}{\ding{52}} & \textcolor{red}{\ding{56}} & \textcolor{red}{\ding{56}} & \textcolor{red}{\ding{56}} \\
\midrule
Fr\'echet Music Distance & \textcolor{ForestGreen}{\ding{52}} & \textcolor{ForestGreen}{\ding{52}} & \textcolor{ForestGreen}{\ding{52}} & \textcolor{red}{\ding{56}} & \textcolor{ForestGreen}{\ding{52}} & \textcolor{ForestGreen}{\ding{52}} \\
Kernel Music Distance & \textcolor{ForestGreen}{\ding{52}} & \textcolor{ForestGreen}{\ding{52}} & \textcolor{ForestGreen}{\ding{52}} & \textcolor{red}{\ding{56}} & \textcolor{ForestGreen}{\ding{52}} & \textcolor{ForestGreen}{\ding{52}} \\
Kernel Performance Distance & \textcolor{ForestGreen}{\ding{52}} & \textcolor{ForestGreen}{\ding{52}} & \textcolor{ForestGreen}{\ding{52}} & \textcolor{ForestGreen}{\ding{52}} & \textcolor{ForestGreen}{\ding{52}} & \textcolor{ForestGreen}{\ding{52}} \\
\midrule
Mahalanobis Distance & \textcolor{ForestGreen}{\ding{52}} & \textcolor{red}{\ding{56}} & \textcolor{ForestGreen}{\ding{52}} & \textcolor{ForestGreen}{\ding{52}} & \textcolor{ForestGreen}{\ding{52}} & \textcolor{ForestGreen}{\ding{52}} \\
Relative Mahalanobis Distance & \textcolor{ForestGreen}{\ding{52}} & \textcolor{red}{\ding{56}} & \textcolor{ForestGreen}{\ding{52}} & \textcolor{ForestGreen}{\ding{52}} & \textcolor{ForestGreen}{\ding{52}} & \textcolor{ForestGreen}{\ding{52}} \\
Marginal Mahalanobis Distance & \textcolor{ForestGreen}{\ding{52}} & \textcolor{red}{\ding{56}} & \textcolor{ForestGreen}{\ding{52}} & \textcolor{red}{\ding{56}} & \textcolor{ForestGreen}{\ding{52}} & \textcolor{ForestGreen}{\ding{52}} \\
\bottomrule
\end{tabular}%
}
\vspace{0.2cm}
\caption{Comparison of attribute-scoped and deep feature metrics for evaluating MIDI piano performances.}
\label{tab:attribute_vs_deep}
\vspace{-5mm}
\end{table}

Inspired by self-supervised learning (SSL), we propose using a pretrained MIDI feature extractor as a perceptual proxy.  Building on Maximum Mean Discrepancy (MMD) \cite{gretton2012kernel, kad}, we introduce two alignment-free, context-aware metrics for symbolic music: Kernel Music Distance (KMD) and Kernel Performance Distance (KPD). KMD compares disjoint sets of musical pieces, while KPD captures differences in performance aspects. Since KMD and KPD are \emph{distributional}, they capture both the fidelity and diversity of samples. We show that the proposed metrics are sensitive to randomness in timing and, unlike attribute-scoped metrics, also respond to contextual corruptions.

Furthermore, we employ different distance measures for \emph{per-sample} ratings, such as the Mahalanobis Distance, that are designed to reflect the robustness of embeddings. We compare two state-of-the-art SSL symbolic music models for understanding solo piano performances: CLaMP3 \cite{wu2025clamp3} and Aria \cite{bradshaw2025scaling}. Our experiments reveal that while CLaMP3 demonstrates robust and competitive rank alignment with human ratings out-of-the-box (Kendall's $\tau$ rank correlation of $0.44$ vs. $0.43 - 0.48$ for individual attributes), the Aria embeddings substantially benefit from post-processing ($0.51$). The pros and cons of each method are summarized in Table \ref{tab:attribute_vs_deep}. We release Pereval, a reproducible evaluation library that implements both attribute-scoped and deep feature metrics for MIDI piano performance datasets.

\section{Background and Related Work}

\subsection{Problem Formulation}

In this section, we describe the problem and provide a brief overview of the field. Given a sequence of note-level score features $y$, we assume an underlying distribution of expressive performance feature sequences $x \in \mathcal{X}$, where each element corresponds to note attributes, such as pitch, velocity, onset, and offset. Performances are typically represented as MIDI events, whereas score features can be stored in MusicXML or MIDI. Suppose that we are given a set of scores $Y = \{ y_i \}_{i=1}^S$, a set of evaluated performances $\tilde{X} = \{ \tilde{x}_{i} \}_{i=1}^M$, and a set of reference performances $X = \{ x_i \}_{i=1}^N$. Our goal is to evaluate the distributional similarity of $\tilde{X}$ with $X$ — for example, comparing beginner interpretations with reference virtuoso references or assessing the realism of generative rendering models.

We omit pedal modeling, as tracking pedals requires a separate time grid. For simplicity, we assume that the sustain pedal extends note durations. In practice, sequence lengths of $x$ and $y$ may differ due to skipped notes, trills, or ornamentation. This suggests that while a performance and its corresponding score should be aligned, such an alignment is not necessarily perfect.

In the context of audio music generation, a line of work proposed to employ deep contextual embeddings \cite{gui2024adapting, kad, huang2025aligning, zhang2025aesthetics}. The study conducted in \cite{Zhang2024FromJudges} revealed that pretrained audio encoders showed promise in performance understanding, yet struggled with subtleties of expert-level interpretations. In \cite{peter2023sounding}, the authors devised expert performance randomization to showcase the limitations of MIDI attribute reconstruction error. 

\subsection{Attribute-scoped Metrics}

Attributes can be aggregated across varying levels of hierarchy within score structures. These include: notes \cite{jeong2019virtuosonet, jeong2019graph, borovik2023scoreperformer, tang2025towards}, onsets \cite{jeong2019virtuosonet, borovik2023scoreperformer, zhang2024dexter}, beats \cite{jeong2019virtuosonet, jeong2019graph, borovik2023scoreperformer}, bars \cite{borovik2023scoreperformer}, and time windows \cite{borovik2023scoreperformer}. We let $\phi_i(x)$ denote the value of the $i$-th attribute unit of a performance $x$. For example, the authors of VirtuosoNet \cite{jeong2019virtuosonet} employ beat-level aggregation to calculate metrics associated with tempo. In this case, $\phi_i(x)$ is the tempo of the $i$-th beat in a performance $x$. For a given score $y$, we denote the set of reference (evaluated) performances derived from $y$ as $X(y)$ ($\tilde{X}(y)$). 

\subsubsection{\textbf{Reconstruction error}} One of the straightforward ways to measure performance quality is to compute reconstruction error relative to a reference set \cite{cancino2018computational, tang2023reconstructing}. However, the error depends on attribute units, making it difficult to aggregate into a single scalar without applying standardization \cite{peter2023sounding}. In generative applications, the reconstruction error is often used to evaluate samples given a style vector extracted from human performances \cite{jeong2019virtuosonet, jeong2019graph, borovik2023scoreperformer}. Another limitation is that it cannot capture sample diversity. Therefore, we focus on correlation-based metrics, which avoid these issues.

\subsubsection{\textbf{Correlation}}The Pearson correlation coefficient is a popular metric that can reveal a linear relationship between attributes \cite{jeong2019virtuosonet, jeong2019graph, zhang2024dexter, tang2025towards, borovik2025symupe} and is arguably the default objective evaluation tool in expressive performance rendering. It is \emph{unit-free}, bounded in $[-1, 1]$, with higher values indicating stronger positive correlation. Each attribute is treated as a random variable, with observations arranged in pairs $\left\{ ( \phi_i(x), \phi_i(\tilde{x}) \right ) \}_{i=1}^{n(x)}$, where $n(x)$ denotes the number of notes. Then, for every pair of performances $x$ and $\tilde{x}$ corresponding to the same score $y$, the Pearson correlation $\rho(\cdot, \cdot)$ is computed independently (i.e., statistics are not shared across different performances). We provide the expression that balances the contribution of each musical piece in the dataset:
\begin{equation}
\small
    \operatorname{corr}(X, \tilde{X} | Y) = \frac{1}{|Y|} \sum_{y \in Y} \frac{\sum\limits_{x \in X(y)} \sum\limits_{\tilde{x} \in \tilde{X}(y)} \rho(\phi(x), \phi(\tilde{x}))}{|X(y)| \cdot |\tilde{X}(y)|}.
\end{equation}

When comparing two different datasets (e.g., human vs. generated performances), the similarity is \emph{inter-}set, measuring average cross-dataset proximity. When comparing samples within the same dataset, the similarity is \emph{intra-}set, reflecting diversity (excluding duplicate pairs where $x = \tilde{x}$).

\subsubsection{\textbf{KL-divergence}} The Kullback-Leibler divergence is used to compare the marginal distributions of attributes \cite{zhang2024dexter, tang2025towards}. In contrast to correlation, KL divergence is alignment-free. It does not compare performances against each other but rather treats each attribute unit without any compositional context.
\begin{equation}
    D_{\operatorname{KL}}(p \| \tilde{p}) = \int p(\phi) \log\left( \frac{p(\phi)}{\tilde{p}(\phi)} \right) d\phi
\end{equation}

In \cite{zhang2024dexter}, the authors employ Gaussian kernels for density estimation and perform Monte Carlo sampling to calculate the KL divergence.

\section{Deep Feature Metrics}\label{sec:deep_metrics}

CLaMP3 supports a diverse range of instruments and multiple modalities, including text, audio, ABC notation, and MIDI \cite{wu2025clamp3}. In contrast, Aria is an autoregressive transformer pretrained on a large-scale corpus of expressive piano performances \cite{bradshaw2025aria, bradshaw2025scaling}. Both models act as mappings $\psi: \mathcal{X} \to \mathbb{R}^d$ for a performance $x \in \mathcal{X}$ that disregard the temporal dimension. This offers two benefits: 1) avoiding the computational burden of note alignment and 2) enabling standard operations like Euclidean distance calculation. To produce global embeddings, both models first divide a MIDI performance into fixed-length chunks. CLaMP3 employs BERT-like encoding for the chunks, followed by average pooling \cite{wu2025clamp3}. Aria embeddings are derived from the last hidden state of each chunk's end-of-sequence token, then averaged across chunks \cite{bradshaw2025scaling}.

\subsection{Fr\'echet Music Distance}
Relevant to our work, \cite{retkowski2024frechet} introduced the Fr\'echet Music Distance (FMD) to evaluate generated symbolic music in ABC and MIDI formats, using embeddings from CLaMP2 \cite{wu2025clamp}. The core assumption behind the Fr\`echet Distance is that the underlying distributions are elliptically contoured, so they are fully described by their mean and covariance matrices. However, embeddings may cluster around distinct musical pieces, violating this assumption. Moreover, it has been shown that the Fr\`echet Distance has a sample-size and model-dependent bias, potentially making model comparisons unreliable \cite{chong2020effectively}.

\subsection{Kernel Music Distance}
Maximum Mean Discrepancy (MMD) is a metric on probability distributions commonly used for two-sample tests \cite{gretton2012kernel}. Several works adopt it as an alternative to the Fr\'echet distance for evaluating generative models, since MMD is unbiased, sample-efficient, and makes no distributional assumptions \cite{ren2016conditional, binkowski2018demystifying, chong2020effectively, jayasumana2024rethinking, kad}. The main practical idea behind MMD is the kernel trick, which is used to calculate the inner product in an infinite-dimensional Reproducing Kernel Hilbert Space. In theory, the mean embeddings in this space can describe an arbitrarily large number of moments of the distributions, provided that the kernel is characteristic. This property ensures that the distributions are equal if and only if their mean embeddings coincide. In this paper, we use a Gaussian kernel on the model embeddings, a common choice for a characteristic kernel:
\begin{equation}
    k(x, x') = \exp\left( -\frac{\| \psi(x) - \psi(x') \|^2}{2 \sigma^2} \right),
\end{equation}
where $\sigma$ is the bandwidth hyperparameter. We set the bandwidth to the median distance between the embeddings in the reference set $X$, which has been shown to be a good baseline \cite{gretton2012kernel, kad}. Then, the unbiased estimate of MMD can be calculated as follows:

{\footnotesize
\begin{multline}
    \widehat{\operatorname{MMD}}^2(X, \tilde{X}) = \frac{1}{N (N-1)} \sum_{i \neq j} k(x_i, x_j) \\ + \frac{1}{M (M-1)} \sum_{i \neq j} k(\tilde{x}_i, \tilde{x}_j)
    - \frac{2}{N M} \sum_{i=1}^N \sum_{j=1}^M k(x_i, \tilde{x}_j),
\end{multline}}
where $N$ and $M$ are sample sizes. The first two terms can be interpreted as the average within-distribution similarities, while the last term indicates cross-dataset similarity. 

Similar to Kernel Audio Distance (KAD), which is an MMD-based metric for audio embeddings \cite{kad}, we coin the rescaled squared MMD for symbolic music as the Kernel Music Distance (KMD):
\begin{equation}
    \operatorname{KMD}(X, \tilde{X}) :=  \alpha \cdot \widehat{\operatorname{MMD}}^2(X, \tilde{X}),
\end{equation}
where $\alpha = 100$ is a global constant factor that is introduced for improved readability.

\subsection{Kernel Performance Distance}
It is not clear whether a naive application of FMD and KMD to performance rendering would be sufficient, as these metrics do not take into account the explicit dependence between the score and performance. For instance, a model could learn the joint distribution of performances quite well but fail to keep them true to the scores. Both FMD and KMD might be highly influenced by popular pieces, since the number of performances per piece can be unbalanced in the dataset. To address these issues, we introduce the Kernel Performance Distance, which is the average per-score KMD or Average MMD (AMMD) \cite{huang2022evaluating}:
\begin{equation}
\small
    \widehat{\operatorname{AMMD}}^2(X, \tilde{X} | Y) = \frac{1}{|Y|} \sum_{y \in Y} \widehat{\operatorname{MMD}}^2(X(y), \tilde{X}(y)).
\end{equation}
Similar to KMD, we employ a Gaussian kernel but use a slightly different heuristic for choosing the bandwidth. Namely, we set the bandwidth to the median \emph{within-score} distance. This ensures that large distances between the performances of different pieces do not affect the kernel values. Finally, as is the case with KMD, we rescale the squared AMMD for readability:
\begin{equation}
    \operatorname{KPD}(X, \tilde{X} | Y) := \alpha \cdot \widehat{\operatorname{AMMD}}^2(X, \tilde{X} | Y)
\end{equation}

\subsection{Per-Sample Pseudo Ratings}
\label{sec:pseudo_ratings}

The proposed deep feature metrics assume an embedding space with a distance function that encodes semantic dissimilarity. Therefore, in addition to distributional similarity, we explore the possibility of devising per-sample pseudo ratings that assess the overall quality and realism of individual performances. We explore different strategies, such as directly plugging KPD or estimating the proximity of a sample to a distribution via the Mahalanobis distance. In relation to our work, similar terminology appears in \cite{gui2024adapting} and \cite{kad} (e.g., "per-song FAD score"). 

\textbf{KPD as a pseudo rating.} For a musical piece with reference embeddings $X$ and a single embedding $\tilde{x}$ (whose rating we want to calculate), we adapt KPD to this setting as the average cross-dataset kernelized distance: $\frac{1}{N} \sum_{i=1}^N k(x_i, \tilde{x})$.

\textbf{Mahalanobis Distance.} Alternatively, we employ the squared Mahalanobis distance between $\psi(\tilde{x})$ and a piece-conditioned distribution with mean $\mu_y$:
\begin{equation}
\label{eq:mahalanobis}
    \textrm{MD}^2(X(y), \tilde{x}) = \left( \psi(\tilde{x}) - \mu_y \right)^{\intercal} \Sigma^{-1} \left( \psi(\tilde{x}) - \mu_y \right).
\end{equation}
Here, $\Sigma$ is a covariance matrix, which is assumed to be shared for each $y \in Y$. This simplification has proven to be effective in class-conditional out-of-distribution (OOD) detection \cite{lee2018simple}, and allows us to alleviate the poor estimation of the covariance matrix for underrepresented pieces. Intuitively, it rescales the contribution of each embedding channel by accounting for correlations within same-piece performances. We compute $\Sigma$ as the weighted average of the score-conditioned covariance matrices $\Sigma_y$, estimated via the Ledoit-Wolf shrinkage algorithm \cite{ledoit2004well}.

\textbf{Relative Mahalanobis Distance.} The embeddings may have non-relevant channels whose contributions oversaturate the distances. Ren et al. \cite{ren2021simple} proposed a modification to Eq. \eqref{eq:mahalanobis} to weaken these contributions by subtracting the squared Mahalanobis distance to the marginal distribution:
\begin{equation}
\label{eq:relative_mahalanobis}
\small
    \textrm{RMD} = \textrm{MD}^2(X(y), \tilde{x}) - \left( \psi(\tilde{x}) - \mu_0 \right)^{\intercal} \Sigma_0^{-1} (\psi(\tilde{x}) - \mu_0),
\end{equation}
where $\mu_0$ and $\Sigma_0$ are piece-agnostic mean and covariance matrix of the reference set, respectively.

\textbf{Marginal Mahalanobis Distance.} Interestingly, the subtracted term in Eq. \eqref{eq:relative_mahalanobis} — the distance to the marginal distribution — can itself serve as a pseudo rating even when \emph{a musical piece is absent in the reference set}. We further explore this use case in our experiments.

\section{Experiments}

\textbf{Data.} We combine the ASAP \cite{peter2023automatic} and ATEPP \cite{zhang2022atepp} datasets of classical piano music, replacing a subset of the original scores with the corresponding files from PDMX \cite{long2025pdmx}. Scores and performances are realigned using Parangonar \cite{peter-offline2023}. Some notes may be missing from the aligned data due to transcription errors, human performance errors, alignment failures, or other factors. We employ linear interpolation of the expressive attributes for these missing notes from the surrounding context. The dataset contains 6263, 698, and 732 performances in the training, validation, and test sets, respectively. The validation and test set contain 46 and 49 distinct musical pieces.

\begin{figure}[t]
  \centering
  \includegraphics[alt={Attribute inter-set correlations (IOI, velocity, and duration) with increasing the number of PianoFlow inference steps, meaning performances become less precise. The intra--set correlations also decrease, indicating the samples become more diverse.},width=0.95\linewidth]{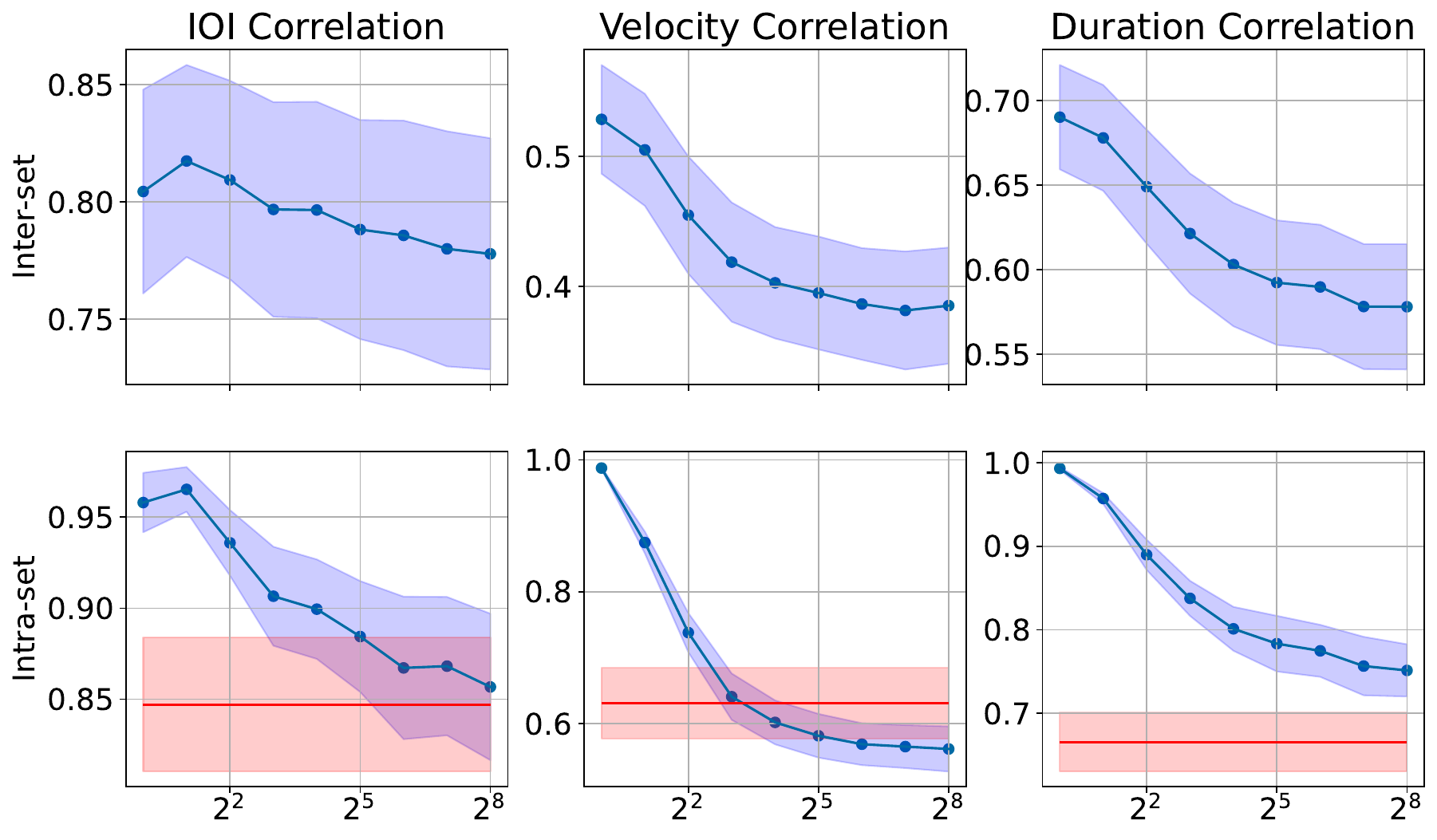}
  \caption{(Top) Inter-set correlations between human and generated performances. (Bottom) The blue lines depict intra-set correlations within generated samples, whereas the red lines shows ground-truth correlations. The x-axis corresponds to the number of function evaluations.}
  \label{fig:correlation}
  \vspace{-5mm}
\end{figure}

\textbf{Evaluated models.} We evaluate three expressive piano rendering models: VirtuosoNet (ISGN), M2M, and PianoFlow. VirtuosoNet is a graph neural network combined with an RNN that learns hierarchical score and performance dependencies \cite{jeong2019virtuosonet, jeong2019graph}. M2M is a BERT-like language model that unmasks tokenized note features in a single step \cite{tang2025towards}. PianoFlow is a flow matching model that treats note features as continuous variables and denoises them in multiple steps \cite{borovik2025symupe}. For VirtuosoNet and M2M, we use the pretrained model checkpoints provided in their respective repositories. We retrained PianoFlow on the training split to simulate hyperparameter selection on the validation set, which is explained in the next subsection.

\begin{figure}[t]
  \centering
  \includegraphics[alt={FMD, KMD, and KPD all exhibit a decreasing trend with the increase of PianoFlow inference steps. This holds true for both CLaMP3 and Aria embeddings.}, width=0.95\linewidth]{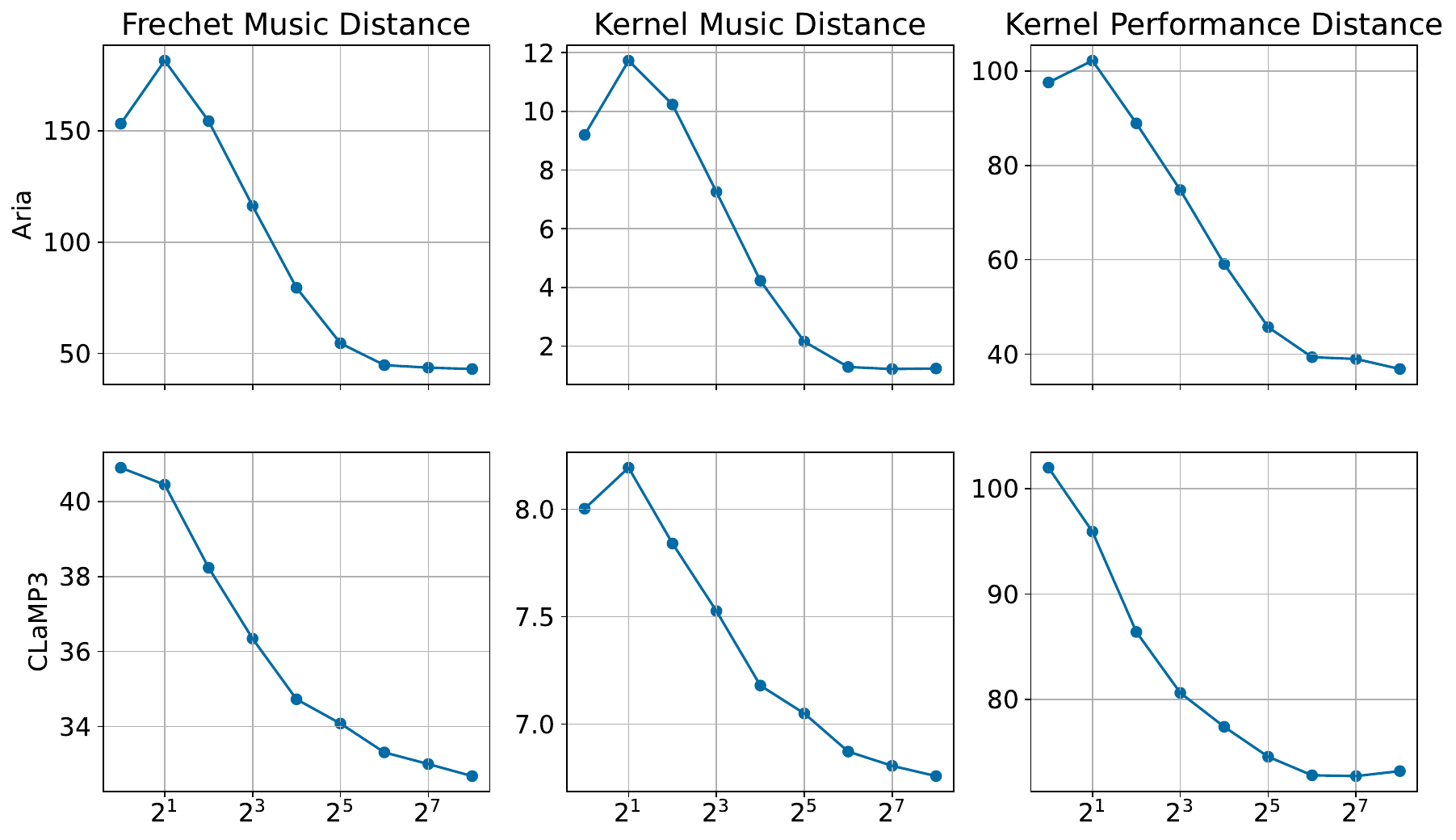}
  \caption{The values of deep feature metrics ($\downarrow$) vs the number of PianoFlow inference steps. In the top and bottom rows, metrics are calculated using Aria and CLaMP3 embeddings, respectively.}
  \label{fig:neural_metrics}
  \vspace{-5mm}
\end{figure}

\subsection{Case Study: Importance of Quantifying Diversity}

Correlation values can be misleading, so both the inter-set and intra-set metrics are necessary to avoid confusion. To demonstrate this, we simulate the selection of a PianoFlow hyperparameter, the number of ODE sampling steps. In real-time applications, practitioners often seek a trade-off between speed and accuracy, where the number of function evaluations (NFE) plays a crucial role.

During inference, we vary the NFE and employ the Euler method to solve the ODE. As shown in the top row of Figure \ref{fig:correlation}, PianoFlow samples become less correlated with real performances as NFE increases. However, the bottom row reveals that single-step PianoFlow exhibits mode collapse, reflected by intra-set correlations close to 1. Increasing NFE mitigates this issue, as generated performances become progressively decorrelated. Ideally, intra-set correlations should match those of the real dataset, although different attributes converge at different rates, making NFE tuning difficult. In contrast, the proposed deep feature metrics exhibit a monotonic decreasing trend (Figure \ref{fig:neural_metrics}). Moreover, they are strongly consistent with each other and match the expected behavior of flow matching models.

\subsection{Sensitivity to Perturbations}

Our next set of experiments analyzes the sensitivity of the embeddings and our proposed deep feature metrics to various synthetic perturbations. 

\begin{figure}[t]
  \centering
  \includegraphics[alt={With the increase of pause intensity and duration, FMD and KMD both increase.}, width=0.70\linewidth]{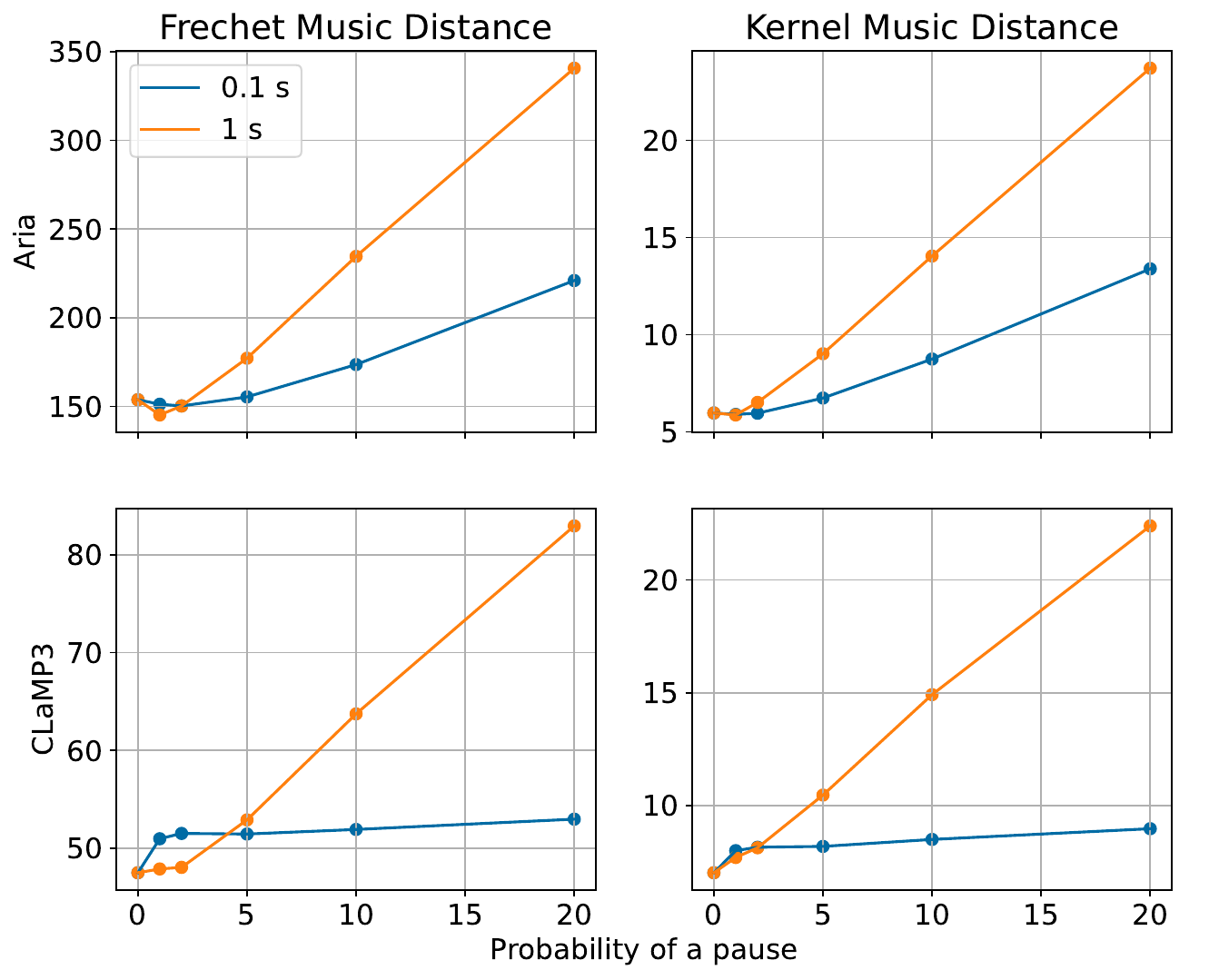}
  \caption{The dependency of FMD and KMD on the probability of synthetic pauses $\eta$ (in percents). The blue and orange lines correspond to the pause durations $t$ of 0.1 and 1 seconds, respectively. The results for Aria are displayed on the top row, and for CLaMP3 on the bottom row.}
  \label{fig:mistakes}
\end{figure}

\textbf{Synthetic pauses.} Humans can identify abrupt changes in tempo, such as sudden pauses in the middle of a musical phrase. These pauses are classified as mistakes and are often perceived negatively, even when they occur rarely in a performance. Motivated by this phenomenon, we perturb the validation set by injecting synthetic mistakes. The level of degradation is controlled by the parameters $t$ and $\eta$. In each performance, we insert a pause of duration $t$ between two consecutive notes with probability $\eta$. Figure \ref{fig:mistakes} demonstrates the discrepancies between the perturbed validation set and the test set, which serves as a reference. Both FMD and KMD, calculated using Aria embeddings, exhibit a linear relationship with the probability of mistakes. Since musical pieces in the validation and test sets do not intersect, this indicates that Aria embeddings capture the temporal regularity of performances, and that deviations from the meter are captured by Aria. This effect is amplified for perturbed sequences with longer pauses, as indicated by the increased steepness of the FMD and KMD lines. The CLaMP3 metrics demonstrate similar behavior for 1-second pauses. However, these embeddings are much less sensitive to shorter pauses, and the metric values do not increase significantly.

\begin{figure}[t]
  \centering
  \includegraphics[alt={Deep feature metrics are almost insensitive to misalignment between performances.}, width=0.95\linewidth]{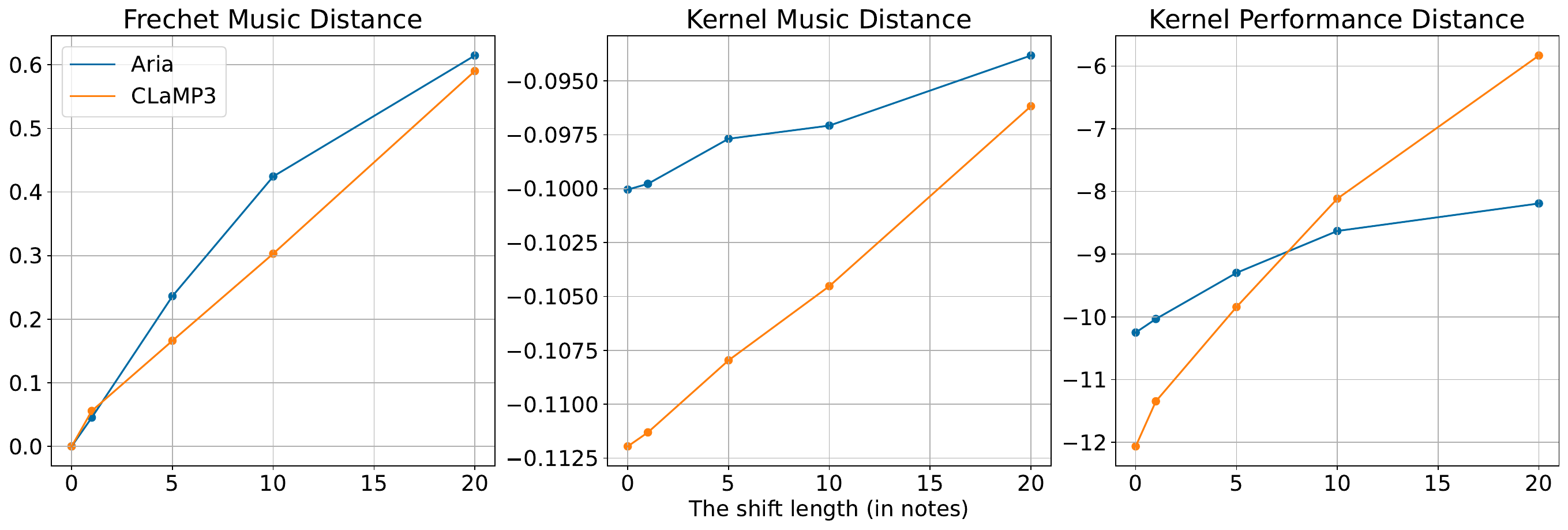}
  \caption{The dependency of deep feature metrics on the note shift. The blue and orange lines correspond to the Aria and CLaMP3 embeddings, respectively.}
  \label{fig:shift}
  \vspace{-5mm}
\end{figure}

\textbf{Sensitivity to alignment.} Human MIDI samples may have small deviations, such as occasional missing or additional notes, trills, and grace notes. These deviations may also stem from inaccuracies in audio-to-MIDI transcription. Therefore, it is quite natural for the note alignment between MIDI piano samples to be imperfect. Since the embeddings are extracted segment-wise and averaged, the corresponding segments for the real and generated performances might be misaligned or shifted with respect to each other. To investigate how such a misalignment affects the proposed deep feature metrics, we simulate it by removing the first few notes from each performance in the validation set. Then, we calculate the discrepancies between the corrupted and original datasets. The results are illustrated in Figure \ref{fig:shift}, where the degree of corruption is controlled by the length of the note shift (from 0 to 20 notes). Despite the monotonic growth, the discrepancies remain considerably low. The presence of negative values in KMD and KPD is due to the unbiasedness of the estimators, indicating that the shifted datasets are nearly indistinguishable from the uncorrupted dataset.

\begin{table}[t]
\centering
\resizebox{0.40\textwidth}{!}{%
\begin{tabular}{@{}lcc@{}}
\toprule
 & Before corruption & After corruption \\
\midrule
IOI Inter-Correlation ($\uparrow$) & $0.87$ & $0.87$ \\
Velocity Inter-Correlation ($\uparrow$) & $0.69$ & $0.68 \pm 0.01$ \\
Duration Inter-Correlation ($\uparrow$) & $0.72$ & $0.72$ \\
\midrule
$\textrm{FMD}_{Aria}$ ($\downarrow$) against validation & $0$ & $13.82 \pm 1.35$ \\
$\textrm{KMD}_{Aria}$ ($\downarrow$) against validation & $-0.11$ & $0.48 \pm 0.10$ \\
$\textrm{KPD}_{Aria}$ ($\downarrow$) against validation & $-11.59$ & $3.37 \pm 1.51$ \\
\midrule
$\textrm{FMD}_{CLaMP3}$ ($\downarrow$) against validation & $0$ & $52.40 \pm 0.47$ \\
$\textrm{KMD}_{CLaMP3}$ ($\downarrow$) against validation & $-0.11$ & $12.46 \pm 0.10$ \\
$\textrm{KPD}_{CLaMP3}$ ($\downarrow$) against validation & $-13.53$ & $78.36 \pm 0.30$ \\
\midrule
$\textrm{FMD}_{Aria}$ ($\downarrow$) against test & $153.77$ & $157.12 \pm 1.09$ \\
$\textrm{KMD}_{Aria}$ ($\downarrow$) against test & $5.97$ & $6.07 \pm 0.08$ \\
\midrule
$\textrm{FMD}_{CLaMP3}$ ($\downarrow$) against test & $47.48$ & $96.15 \pm 0.36$ \\
$\textrm{KMD}_{CLaMP3}$ ($\downarrow$) against test & $7.03$ & $21.68 \pm 0.11$ \\
\bottomrule
\end{tabular}%
}
\vspace{0.2cm}
\caption{The effect of distorting the dataset by transferring velocities. The metrics are reported against different reference sets for the validation set before and after corruption. The metrics for corrupted datasets are averaged across 5 independent random perturbations.}
\vspace{-5mm}
\label{tab:transfer_velocities}
\end{table}

\textbf{Contextual perturbations.} Attribute-scoped metrics account only for a single facet of a performance. Deep feature metrics, in contrast, have the potential to capture contextual dependencies beyond traditional metrics. In our next synthetic perturbation experiment, we transfer note velocities from one performance to another. That is, each sample in the dataset remains unchanged except for the velocities, which are taken from another interpretation. This preserves dataset-level statistics, such as Pearson correlation and KL-Divergence, because the attribute sequences are fixed. Yet the original and perturbed performances should sound different, since artistic choices behind dynamics affect tempo and articulation, and vice versa. 

To measure this effect, we perturb the validation set and compare it with both the validation and test sets. Table \ref{tab:transfer_velocities} shows how this corruption affects the metrics. As expected, correlation fails to capture these perturbations, where a small change in the velocity inter-correlation can be explained by imperfect note alignment. However, deep feature metrics detect the distributional shift consistently with respect to both reference sets. Although the changes in metrics computed with the Aria embeddings are relatively moderate, such changes are significantly more pronounced with the CLaMP3 embeddings. We note that such dramatic increase in the case of CLaMP3 does not necessarily indicate an advantage over Aria. The listening study conducted by Peter et al. \cite{peter2023sounding} revealed that humans are less sensitive to randomness in dynamics than in tempo or articulation. Nevertheless, the velocity-corrupted dataset exhibits a range of $\textrm{KMD}_{CLaMP3}$ values similar to that of the dataset with synthetic pauses (Figure \ref{fig:mistakes}).

\subsection{Agreement with Human Perception}

To investigate how well SSL embeddings align with human perception, we conducted an adaptive online listening test comprising 29 musical pieces from the test set. The test was implemented with PsyNet \cite{harrison2020psynet} and divided into 10 rounds, where participants rated the naturalness and expression on a 7-point Likert scale. The performances were presented as 15-second excerpts in random order, rendered using the Grand Piano preset from the Ableton digital audio workstation. The listening test involved five models: M2M, VirtuosoNet, and PianoFlow with 2, 16, and 128 ODE steps. We recruited 23 participants, and the collected user data yielded 810 human ratings.

\begin{table}[t]
\centering
\resizebox{0.4\textwidth}{!}{%
\begin{tabular}{@{}lcc@{}}
\toprule
& \textbf{Naturalness} & \textbf{Expression} \\
\midrule
IOI Correlation & $0.45 \pm 0.10$ & $0.35 \pm 0.12$ \\
Velocity Correlation & $0.43 \pm 0.13$ & $0.37 \pm 0.14$ \\
Duration Correlation & $\underline{0.48} \pm 0.10$ & $0.40 \pm 0.13$ \\
Aggregated Correlation & $\mathbf{0.51} \pm 0.09$ & $\underline{0.43} \pm 0.12$ \\
\midrule
Kernel Perf. Distance (CLaMP3) & $0.44 \pm 0.12$ & $0.36 \pm 0.14$ \\
Kernel Perf. Distance (Aria) & $0.29 \pm 0.14$ & $0.25 \pm 0.16$ \\
\midrule
Mahalanobis (CLaMP3) & $0.42 \pm 0.12$ & $0.36 \pm 0.12$ \\
Relative Mahalanobis (CLaMP3) & $0.38 \pm 0.13$ & $0.36 \pm 0.13$ \\
Mahalanobis (Aria) & $0.34 \pm 0.13$ & $0.30 \pm 0.15$ \\
Relative Mahalanobis (Aria) & $\mathbf{0.51} \pm 0.13$ & $\mathbf{0.45} \pm 0.10$ \\
\midrule
Marginal Mahalanobis (CLaMP3) & $0.42 \pm 0.11$ & $0.36 \pm 0.13$ \\
Marginal Mahalanobis (Aria) & $0.24 \pm 0.14$ & $0.22 \pm 0.16$ \\
\bottomrule
\end{tabular}%
}
\vspace{0.2cm}
\caption{Average Kendall-$\tau_B$ rank correlation between the pseudo ratings and human Mean Opinion Scores across the scores. "$\pm$" indicates the 95\% confidence intervals. The best and second best values are highlighted in bold and underscored, respectively.}
\label{tab:perception_alignment}
\end{table}

\begin{figure}[t]
  \centering
  \includegraphics[alt={Agreement of aggregated correlation and Relative Mahalanobis Distance (Aria) with human ratings. Both pseudo ratings exhibit monotonic agreement with human judgements.}, width=0.95\linewidth]{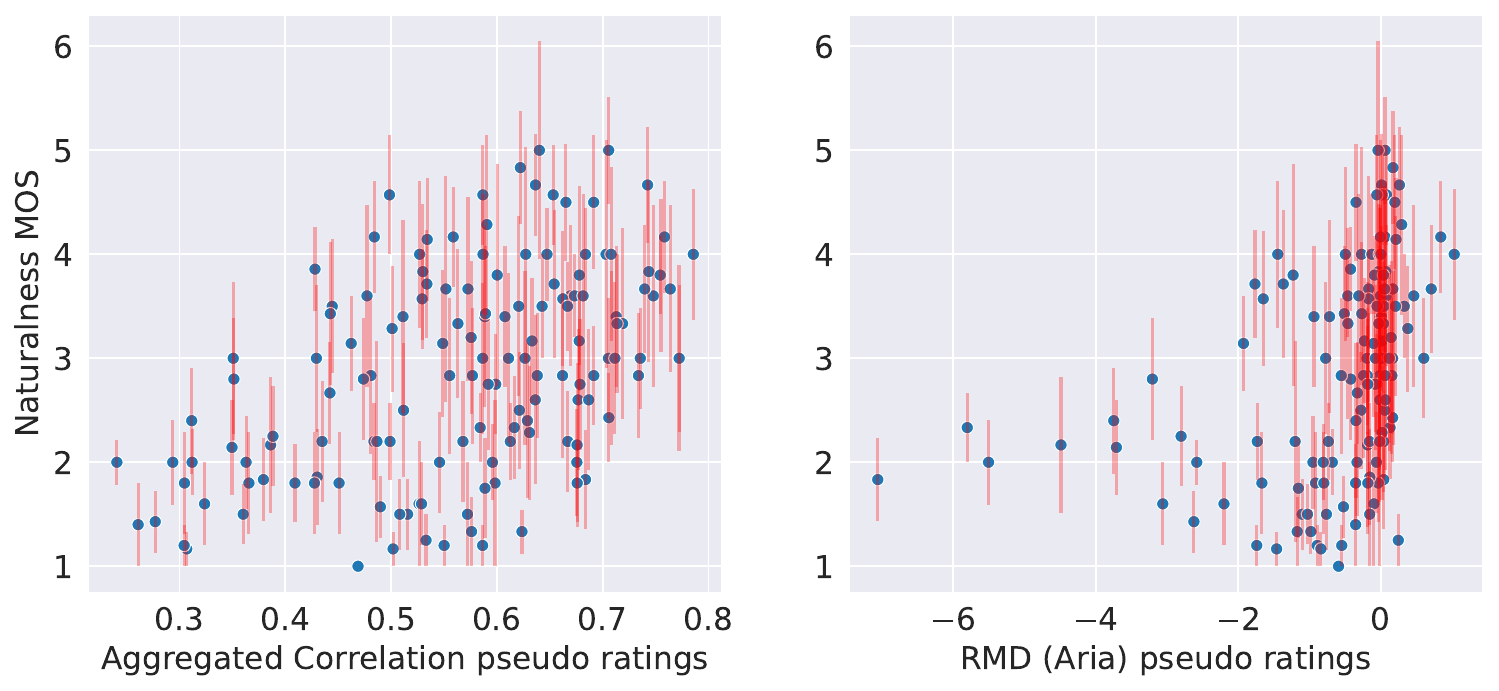}
  \caption{Pseudo ratings (x-axis) vs human MOS (y-axis). Each point corresponds to a generated performance. Red lines indicate standard error bars.}
  \label{fig:pseudo_ratings}
  \vspace{-5mm}
\end{figure}

We compared the pseudo ratings proposed in Section \ref{sec:pseudo_ratings} with attribute-scoped ratings based on the average inter-onset interval (IOI), velocity, and duration correlations. In addition, we introduced \emph{aggregated correlation} that combined these ratings with equal weights. Each of the 145 generated performances in the listening test was assigned a Mean Opinion Score (MOS) by averaging human ratings. We measured the agreement of pseudo ratings with MOS for both naturalness and expression using Kendall's $\tau_B$ correlation coefficient \cite{kendall1945treatment}, whose values fall within the range $[-1, 1]$. To ensure consistent comparison between methods, we flipped the sign of distance-based ratings to convert them into similarity measures. In Table \ref{tab:perception_alignment}, we report the average per-score Kendall $\tau_B$ for each of the methods. The devised ratings showed positive alignment with the human ratings across both naturalness and expression dimensions. Using kernelized distances, CLaMP3 outperformed Aria by a significant margin and was competitive with attribute-scoped correlations. However, Aria embeddings benefited from employing the Mahalanobis distance, with RMD matching the strong performance of aggregated correlation. For Marginal Mahalanobis Distance, we estimated the covariance matrix on the training set, simulating a scenario where the reference set did not contain pieces from the listening test. In this case, CLaMP3 further exhibited robustness, with only a slight drop in agreement with respect to naturalness. Example scatter plots for RMD and aggregated correlation are provided in Figure \ref{fig:pseudo_ratings}.

\begin{table}[t]
\centering
\resizebox{0.45\textwidth}{!}{%
\begin{tabular}{@{}lccccc@{}}
\toprule
        & M2M & VirtuosoNet & PianoFlow-128 & PianoFlow-16 & PianoFlow-2 \\
\midrule
Naturalness ($\uparrow$) & \cellcolor{color_a}1.79 & \cellcolor{color_b}2.88 & \cellcolor{color_c}3.12 & \cellcolor{color_d}3.24 & \cellcolor{color_e}3.55 \\
Expression ($\uparrow$)  & \cellcolor{color_a}2.30 & \cellcolor{color_b}3.07 & \cellcolor{color_c}3.35  & \cellcolor{color_d}3.51 & \cellcolor{color_e}3.63 \\
\midrule
\midrule
IOI KLD ($\downarrow$)      & \cellcolor{color_a}0.034 & \cellcolor{color_d}0.016 & \cellcolor{color_e}0.010 & \cellcolor{color_c}0.022 & \cellcolor{color_b}0.027  \\
Velocity KLD ($\downarrow$) & \cellcolor{color_d}0.023 & \cellcolor{color_a}0.131 & \cellcolor{color_e}0.021 & \cellcolor{color_c}0.028 & \cellcolor{color_b}0.079 \\
Duration KLD ($\downarrow$) & \cellcolor{color_a}0.221 & \cellcolor{color_c}0.027 & \cellcolor{color_e}0.002 & \cellcolor{color_d}0.011 & \cellcolor{color_b}0.090\\
\midrule
IOI Inter-Corr. ($\uparrow$)      & \cellcolor{color_a}0.68 & \cellcolor{color_c}0.82 & \cellcolor{color_c}0.82 & \cellcolor{color_c}0.82 & \cellcolor{color_e}0.84 \\
Velocity Inter-Corr. ($\uparrow$) & \cellcolor{color_a}0.25 & \cellcolor{color_d}0.55 & \cellcolor{color_b}0.49 & \cellcolor{color_c}0.51 & \cellcolor{color_e}0.60\\
Duration Inter-Corr. ($\uparrow$) & \cellcolor{color_b}0.36 & \cellcolor{color_a}0.30 & \cellcolor{color_c}0.49 & \cellcolor{color_d}0.50 & \cellcolor{color_e}0.58 \\
\midrule
IOI Intra-Corr. ($\downarrow$)      & \cellcolor{color_e}0.86 & \cellcolor{color_a}0.97 & \cellcolor{color_d}0.92 & \cellcolor{color_c}0.93 & \cellcolor{color_a}0.97 \\
Velocity Intra-Corr. ($\downarrow$) & \cellcolor{color_e}0.67 & \cellcolor{color_a}0.94 & \cellcolor{color_d}0.71 & \cellcolor{color_c}0.74 & \cellcolor{color_b}0.93 \\
Duration Intra-Corr. ($\downarrow$) & \cellcolor{color_e}0.68 & \cellcolor{color_c}0.75 & \cellcolor{color_c}0.75 & \cellcolor{color_b}0.79 & \cellcolor{color_a}0.95 \\
\midrule
$\operatorname{FMD}_{Aria}$ ($\downarrow$) & \cellcolor{color_a}302   & \cellcolor{color_b}236   & \cellcolor{color_e}37   & \cellcolor{color_d}72    & \cellcolor{color_c}173   \\
$\operatorname{KMD}_{Aria}$ ($\downarrow$) & \cellcolor{color_a}24.8  & \cellcolor{color_b}17.7  & \cellcolor{color_e}1.4  & \cellcolor{color_d}4.5   & \cellcolor{color_c}13.2  \\
$\operatorname{KPD}_{Aria}$ ($\downarrow$) & \cellcolor{color_a}137   & \cellcolor{color_b}120   & \cellcolor{color_e}45   & \cellcolor{color_d}69    & \cellcolor{color_c}113 \\
\midrule
$\operatorname{FMD}_{CLaMP3}$ ($\downarrow$) & \cellcolor{color_a}72   & \cellcolor{color_b}51   & \cellcolor{color_e}30   & \cellcolor{color_d}32   & \cellcolor{color_c}37 \\
$\operatorname{KMD}_{CLaMP3}$ ($\downarrow$) & \cellcolor{color_a}24.9 & \cellcolor{color_b}17.1 & \cellcolor{color_e}9.7  & \cellcolor{color_d}10.3 & \cellcolor{color_c}11.5 \\
$\operatorname{KPD}_{CLaMP3}$ ($\downarrow$) & \cellcolor{color_a}125  & \cellcolor{color_b}114   & \cellcolor{color_e}88   & \cellcolor{color_d}92   & \cellcolor{color_c}107 \\
\bottomrule
\end{tabular}%
}
\vspace{0.2cm}
\caption{The results of the listening test and the metrics values for the models compared against the test set. The models (columns) are sorted according to the human ratings, with the purple-to-green color scheme indicating their respective rankings.}
\label{tab:metrics}
\vspace{-5mm}
\end{table}

Table \ref{tab:metrics} presents the results of the listening test and metric values calculated for each model. Naturalness and expression shared similar ranking patterns, with the latter rated slightly higher by the participants. Across attribute-scoped metrics, IOI inter-correlation achieved perfect agreement with human rankings. Nevertheless, an evaluation based on different attributes led to different rankings for both KL divergence and correlation. Deep feature metrics consistently produced the same ranking that coincided with human perception, with the exception of the variants of PianoFlow, which can be explained by different levels of diversity.

\section{Conclusion}

In this paper, we discuss the limitations of attribute-scoped metrics for evaluating expressive MIDI performances. Inspired by recent advances in adjacent fields, we propose deep feature metrics based on rich embeddings from symbolic music understanding models. Our experiments demonstrate the agreement of these metrics with human perception, with CLaMP3 showing competitive performance with Pearson correlation on individual attributes. Notably, CLaMP3 also excels in a \emph{reference-free} scenario, where listening test samples are absent in the reference set. Meanwhile, Aria's moderate agreement with human ratings improves when using the Mahalanobis Distance, and employing RMD enables it to be on par with aggregated correlation. We hypothesize that this improvement occurs because Aria was trained on augmented views of MIDI samples, which degraded its ability to discriminate between performances. We hope these contributions will facilitate new directions for evaluating expressive piano performances.

\section{Ethics Statement}
Participation in our listening test was voluntary, and all participants consented to taking the test. The call for participation was posted across online classical music communities. All collected data were anonymous, as no personal data were collected at any stage. In this work, the datasets used in training and evaluating models are publicly available under a CC BY-NC-SA 4.0 license. The scope of our work is limited to the Western classical solo piano repertoire; we would be interested to extend this to other cultures and genres. We acknowledge that human piano interpretations are complex, rich, and multifaceted objects that are difficult to interpret through distributional and per-sample metrics.

\bibliography{ISMIRtemplate}

@inproceedings{bradshaw2025scaling,
  title={Scaling Self-Supervised Representation Learning for Symbolic Piano Performance},
  author={Bradshaw, Louis and Spangher, Alexander and Fan, Honglu and Biderman, Stella and Colton, Simon},
  booktitle={Proceedings of the International Society for Music Information Retrieval Conference (ISMIR)},
  year={2025}
}

@inproceedings{bradshaw2025aria,
  title={Aria-MIDI: A Dataset of Piano MIDI Files for Symbolic Music Modeling},
  author={Bradshaw, Louis and Colton, Simon},
  booktitle={International Conference on Learning Representations},
  year={2025},
  url={https://openreview.net/forum?id=X5hrhgndxW}, 
}

@article{cancino2017evaluation,
  title={An evaluation of linear and non-linear models of expressive dynamics in classical piano and symphonic music},
  author={Cancino-Chac{\'o}n, Carlos Eduardo and Gadermaier, Thassilo and Widmer, Gerhard and Grachten, Maarten},
  journal={Machine Learning},
  volume={106},
  number={6},
  pages={887--909},
  year={2017},
  publisher={Springer}
}

@inproceedings{jeong2019graph,
  title={Graph neural network for music score data and modeling expressive piano performance},
  author={Jeong, Dasaem and Kwon, Taegyun and Kim, Yoojin and Nam, Juhan},
  booktitle={International conference on machine learning},
  pages={3060--3070},
  year={2019},
  organization={PMLR}
}

@inproceedings{borovik2023scoreperformer,
  title={ScorePerformer: Expressive Piano Performance Rendering With Fine-Grained Control.},
  author={Borovik, Ilya and Viro, Vladimir},
  booktitle={ISMIR},
  pages={588--596},
  year={2023}
}

@inproceedings{tang2025towards,
  title={Towards an integrated approach for expressive piano performance synthesis from music scores},
  author={Tang, Jingjing and Cooper, Erica and Wang, Xin and Yamagishi, Junichi and Fazekas, Gy{\"o}rgy},
  booktitle={ICASSP 2025-2025 IEEE International Conference on Acoustics, Speech and Signal Processing (ICASSP)},
  pages={1--5},
  year={2025},
  organization={IEEE}
}

@article{zhang2024dexter,
  title={Dexter: Learning and controlling performance expression with diffusion models},
  author={Zhang, Huan and Chowdhury, Shreyan and Cancino-Chac{\'o}n, Carlos Eduardo and Liang, Jinhua and Dixon, Simon and Widmer, Gerhard},
  journal={Applied Sciences},
  volume={14},
  number={15},
  pages={6543},
  year={2024},
  publisher={MDPI}
}

@inproceedings{jeong2019virtuosonet,
  title={VirtuosoNet: A Hierarchical RNN-based System for Modeling Expressive Piano Performance.},
  author={Jeong, Dasaem and Kwon, Taegyun and Kim, Yoojin and Lee, Kyogu and Nam, Juhan},
  booktitle={ISMIR},
  pages={908--915},
  year={2019}
}

@inproceedings{gui2024adapting,
  title={Adapting frechet audio distance for generative music evaluation},
  author={Gui, Azalea and Gamper, Hannes and Braun, Sebastian and Emmanouilidou, Dimitra},
  booktitle={ICASSP 2024-2024 IEEE International Conference on Acoustics, Speech and Signal Processing (ICASSP)},
  pages={1331--1335},
  year={2024},
  organization={IEEE}
}

@article{retkowski2024frechet,
  title={Frechet music distance: A metric for generative symbolic music evaluation},
  author={Retkowski, Jan and St{\k{e}}pniak, Jakub and Modrzejewski, Mateusz},
  journal={arXiv preprint arXiv:2412.07948},
  year={2024}
}

@article{gretton2012kernel,
  title={A kernel two-sample test},
  author={Gretton, Arthur and Borgwardt, Karsten M and Rasch, Malte J and Sch{\"o}lkopf, Bernhard and Smola, Alexander},
  journal={The journal of machine learning research},
  volume={13},
  number={1},
  pages={723--773},
  year={2012},
  publisher={JMLR. org}
}

@inproceedings{wu2025clamp,
  title={CLaMP 2: Multimodal Music Information Retrieval Across 101 Languages Using Large Language Models},
  author={Wu, Shangda and Wang, Yashan and Yuan, Ruibin and Zhancheng, Guo and Tan, Xu and Zhang, Ge and Zhou, Monan and Chen, Jing and Mu, Xuefeng and Gao, Yuejie and others},
  booktitle={Findings of the Association for Computational Linguistics: NAACL 2025},
  pages={435--451},
  year={2025}
}

@inproceedings{wu2025clamp3,
  title={Clamp 3: Universal music information retrieval across unaligned modalities and unseen languages},
  author={Wu, Shangda and Zhancheng, Guo and Yuan, Ruibin and Jiang, Junyan and Doh, Seungheon and Xia, Gus and Nam, Juhan and Li, Xiaobing and Yu, Feng and Sun, Maosong},
  booktitle={Findings of the Association for Computational Linguistics: ACL 2025},
  pages={2605--2625},
  year={2025}
}

@article{peter2023automatic,
  title={Automatic note-level score-to-performance alignments in the ASAP dataset},
  author={Peter, Silvan David and Cancino-Chac{\'o}n, Carlos Eduardo and Foscarin, Francesco and McLeod, Andrew Philip and Henkel, Florian and Karystinaios, Emmanouil and Widmer, Gerhard},
  journal={Transactions of the International Society for Music Information Retrieval},
  volume={6},
  number={1},
  year={2023}
}

@inproceedings{zhang2022atepp,
  title={ATEPP: A Dataset of Automatically Transcribed Expressive Piano Performance},
  author={Zhang, Huan and Tang, Jingjing and Rafee, Syed RM and Dixon, Simon and Fazekas, George and Wiggins, Geraint A},
  booktitle={ISMIR 2022 Hybrid Conference},
  year={2022}
}

@inproceedings{long2025pdmx,
  title={PDMX: A Large-Scale Public Domain MusicXML Dataset for Symbolic Music Processing},
  author={Long, Phillip and Novack, Zachary and Berg-Kirkpatrick, Taylor and McAuley, Julian},
  booktitle={ICASSP 2025-2025 IEEE International Conference on Acoustics, Speech and Signal Processing (ICASSP)},
  pages={1--5},
  year={2025},
  organization={IEEE}
}

@inproceedings{peter-offline2023,
  title={Online Symbolic Music Alignment with Offline Reinforcement Learning},
  author={Peter, Silvan David},
  booktitle={International Society for Music Information Retrieval Conference {(ISMIR)}},
  year={2023}
}

@inproceedings{tang2023reconstructing,
  title={Reconstructing Human Expressiveness in Piano Performances with a Transformer Network},
  author={Tang, Jingjing and Wiggins, Geraint and Fazekas, Gy{\"o}rgy},
  booktitle={International Symposium on Computer Music Multidisciplinary Research},
  pages={83--96},
  year={2023}
}

@inproceedings{borovik2025symupe,
  title={SyMuPe: Affective and Controllable Symbolic Music Performance},
  author={Borovik, Ilya and Gavrilev, Dmitrii and Viro, Vladimir},
  booktitle={Proceedings of the 33rd ACM International Conference on Multimedia},
  pages={10699--10708},
  year={2025}
}

@article{cancino2018computational,
  title={Computational modeling of expressive music performance with linear and non-linear basis function models},
  author={Cancino-Chac{\'o}n, Carlos Eduardo},
  journal={Johannes Kepler University, Linz},
  year={2018}
}

@inproceedings{harrison2020psynet,
  title     = {Gibbs Sampling with People},
  booktitle = {Advances in Neural Information Processing Systems},
  author    = {Harrison, Peter M. C. and Marjieh, Raja and Adolfi, Federico and
               {van Rijn}, Pol and Anglada-Tort, Manuel and Tchernichovski, Ofer and
               Larrouy-Maestri, Pauline and Jacoby, Nori},
  date      = {2020},
  volume    = {33},
  url       = {https://arxiv.org/abs/2008.02595}
}

@article{kad,
    author={Chung, Yoonjin and Eu, Pilsun and Lee, Junwon and Choi, Keunwoo and Nam, Juhan and Chon, Ben Sangbae},
    title={KAD: No More FAD! An Effective and Efficient Evaluation Metric for Audio Generation}, 
    journal = {arXiv:2502.15602},
    url = {https://arxiv.org/abs/2502.15602},
    year = {2025}
}

@inproceedings{jayasumana2024rethinking,
  title={Rethinking fid: Towards a better evaluation metric for image generation},
  author={Jayasumana, Sadeep and Ramalingam, Srikumar and Veit, Andreas and Glasner, Daniel and Chakrabarti, Ayan and Kumar, Sanjiv},
  booktitle={Proceedings of the IEEE/CVF Conference on Computer Vision and Pattern Recognition},
  pages={9307--9315},
  year={2024}
}

@inproceedings{chong2020effectively,
  title={Effectively unbiased fid and inception score and where to find them},
  author={Chong, Min Jin and Forsyth, David},
  booktitle={Proceedings of the IEEE/CVF conference on computer vision and pattern recognition},
  pages={6070--6079},
  year={2020}
}

@article{huang2022evaluating,
  title={Evaluating aleatoric uncertainty via conditional generative models},
  author={Huang, Ziyi and Lam, Henry and Zhang, Haofeng},
  journal={arXiv preprint arXiv:2206.04287},
  year={2022}
}

@article{kendall1945treatment,
  title={The treatment of ties in ranking problems},
  author={Kendall, Maurice G},
  journal={Biometrika},
  volume={33},
  number={3},
  pages={239--251},
  year={1945},
  publisher={JSTOR}
}

@inproceedings{peter2023sounding,
  title={Sounding out reconstruction error-based evaluation of generative models of expressive performance},
  author={Peter, Silvan David and Cancino-Chac{\'o}n, Carlos Eduardo and Karystinaios, Emmanouil and Widmer, Gerhard},
  booktitle={Proceedings of the 10th International Conference on Digital Libraries for Musicology},
  pages={58--66},
  year={2023}
}

@article{lee2018simple,
  title={A simple unified framework for detecting out-of-distribution samples and adversarial attacks},
  author={Lee, Kimin and Lee, Kibok and Lee, Honglak and Shin, Jinwoo},
  journal={Advances in neural information processing systems},
  volume={31},
  year={2018}
}

@article{ledoit2004well,
  title={A well-conditioned estimator for large-dimensional covariance matrices},
  author={Ledoit, Olivier and Wolf, Michael},
  journal={Journal of multivariate analysis},
  volume={88},
  number={2},
  pages={365--411},
  year={2004},
  publisher={Elsevier}
}

@article{ren2021simple,
  title={A simple fix to mahalanobis distance for improving near-ood detection},
  author={Ren, Jie and Fort, Stanislav and Liu, Jeremiah and Roy, Abhijit Guha and Padhy, Shreyas and Lakshminarayanan, Balaji},
  journal={arXiv preprint arXiv:2106.09022},
  year={2021}
}

@inproceedings{huang2025aligning,
  title={Aligning Text-to-Music Evaluation with Human Preferences},
  author={Huang, Yichen and Novack, Zachary and Saito, Koichi and Shi, Jiatong and Watanabe, Shinji and Mitsufuji, Yuki and Thickstun, John and Donahue, Chris},
  booktitle={Proceedings of the International Society for Music Information Retrieval Conference (ISMIR)},
  year={2025}
}

@inproceedings{Zhang2024FromJudges,
    address = {San Francisco, USA},
    author = {Zhang, Huan and Liang, Jinhua and Dixon, Simon},
    booktitle = {Proceedings of the International Society for Music Information Retrieval Conference (ISMIR)},
    title = {From Audio Encoders to Piano Judges: Benchmarking Performance Understanding for Solo Piano},
    year = {2024}
}

@INPROCEEDINGS{Zhang2025LLaQo,
  author={Zhang, Huan and Cheung, Vincent K.M. and Nishioka, Hayato and Dixon, Simon and Furuya, Shinichi},
  booktitle={ICASSP 2025 - 2025 IEEE International Conference on Acoustics, Speech and Signal Processing (ICASSP)}, 
  title={LLaQo: Towards a Query-Based Coach in Expressive Music Performance Assessment}, 
  year={2025},
  pages={1-5},
  doi={10.1109/ICASSP49660.2025.10890522}}

@inproceedings{zhang2025aesthetics,
  title={From aesthetics to human preferences: Comparative perspectives of evaluating text-to-music systems},
  author={Zhang, Huan and Liang, Jinhua and Phan, Huy and Wang, Wenwu and Benetos, Emmanouil},
  booktitle={2025 IEEE 35th International Workshop on Machine Learning for Signal Processing (MLSP)},
  pages={1--6},
  year={2025},
  organization={IEEE}
}

@inproceedings{binkowski2018demystifying,
  author       = {Mikolaj Binkowski and
                  Danica J. Sutherland and
                  Michael Arbel and
                  Arthur Gretton},
  title        = {Demystifying {MMD} GANs},
  booktitle    = {6th International Conference on Learning Representations, {ICLR} 2018,
                  Vancouver, BC, Canada, April 30 - May 3, 2018, Conference Track Proceedings},
  publisher    = {OpenReview.net},
  year         = {2018},
  url          = {https://openreview.net/forum?id=r1lUOzWCW},
  bibsource    = {dblp computer science bibliography, https://dblp.org}
}

@article{ren2016conditional,
  title={Conditional generative moment-matching networks},
  author={Ren, Yong and Zhu, Jun and Li, Jialian and Luo, Yucen},
  journal={Advances in Neural Information Processing Systems},
  volume={29},
  year={2016}
}

%
%
%
%

\end{document}